\documentclass{pasj00}

\begin{document}
\SetRunningHead{M. Shimojo}{Unusual migration of the prominence activities}
\Received{2013/05/13}
\Accepted{2013/06/08}

\title{Unusual migration of the prominence activities \\ in the southern hemisphere during Cycle 23--24  }

\author{Masumi \textsc{Shimojo}}
\affil{National Astronomical Observatory of Japan, \\Mitaka, Tokyo, 183-1305, Japan}
\email{masumi.shimojo@nao.ac.jp}

\KeyWords{Sun: prominences --- Sun: magnetic fields --- Sun: radio radiation} 

\maketitle

\begin{abstract}
The solar activity in Cycle 23--24 shows differences from the previous cycles that were observed with modern instruments, e.g. 
long cycle duration and a small number of sunspots. To appreciate the anomalies further, we investigated the prominence eruptions and 
disappearances observed with the Nobeyama Radioheliograph during over 20 years. Consequently, we found that the 
occurrence of the prominence activities in the northern hemisphere is normal because the period of the number variation is 11 years 
and the migration of the producing region of the prominence activities traces the migration of 11 years ago. 
On the other hand, the migration in the southern hemisphere significantly differs from that in the northern hemisphere and the previous cycles. 
The prominence activities occurred over -50 degrees latitude in spite of the late decay phase of Cycle 23, and the number of the 
prominence activities in the higher latitude region (over -65 degrees) is very small even near the solar maximum of Cycle 24. 
The results suggest that the anomalies of the global magnetic field distribution started at the solar maximum of Cycle 23. 
Comparison of the butterfly diagram of the prominence activities with the magnetic butterfly diagram indicates that the timing of \textquotedblleft the 
rush to the pole" and the polar magnetic field closely relates to the unusual migration. Considering that the rush to the pole is made of 
the sunspots, the hemispheric asymmetry of the sunspots and the strength of the polar magnetic fields are essential for understanding 
the anomalies of the prominence activities.
\end{abstract}

\section{Introduction}

The solar activity of the recent solar cycles, Cycle 23--24, shows the significant differences from the previous cycles that were observed over the course of half of a century with modern instruments such as a magnetograph. The noticeable differences are the long cycle duration of Cycle 23 
(over 12 years) and the low activity of Cycle 24. Many authors have already reported the other differences, for examples, the decreasing in 
magnetic field strength in the sunspots \citep{Livi12}, the strong hemispheric asymmetry of the magnetic flux \citep{Petr12}, the weakness of the 
polar fields of Cycle 23 \citep{Wang09,Gopa12}, the delay of the polarity reversal in the south polar region at Cycle 24 \citep{Shio12,Sval13} 
and the large and numerous low-latitude coronal holes \citep{Lee09}. Naturally, the differences have impacts on the heliosphere. 
The asymmetry of the streamer and the heliospheric current sheet, the reduction of the solar-wind mass flux, the weakness and the 
low-latitude non-uniformity of the interplanetary mean field were reported \citep{Wang09,Thom11}. To understand the anomalies of 
the solar activity in the recent solar cycles, we need to investigate the global distribution and evolution of the photospheric magnetic field. 
One way is to investigate directly the magnetic field distribution using the magnetograms. However, it is relatively difficult to 
understand an outline of the global magnetic field distribution because the magnetograms include the numerous fine structures. 
Dark filaments were used for understanding the global magnetic field distribution because a dark filament always lies on a magnetic neutral line. 
The Solar-Geophysical Data (SGD) published by the NOAA National Geophysical Data Center had carried the Carrington map that indicates 
the locations of dark filaments until the 1990s for this purpose.

The Nobeyama Radioheliograph (NoRH) is an interferometer in the microwave range, 
and can observe the thermal emission from a prominence well.  Based on the advantage of the NoRH for prominence observations, 
\citet{Shim06} developed a semi-automatic detection method of prominence eruptions and disappearances for the NoRH data, and made a 
butterfly diagram of the prominence eruptions and disappearances. Considering that all dark filaments disappear by erupting and/or falling down,
the distribution of the prominence eruptions and disappearances indicate that of the dark filaments. For this reason, the prominence 
eruption and disappearance are not distinguished and they are called \textquotedblleft prominence activity" in this paper.  To understand the global distribution and 
evolution of magnetic field of the Sun, we applied the prominence activity detection method developed by \citet{Shim06} to over 20 years of NoRH 
data and investigated the migration of the producing region of the prominence activities. 

In the next section, we briefly describe the observing instruments, the result of the detection, and the limitation of the detection method related to this 
study. In Section 3, we show the solar cycle dependence of the occurrence of prominence activities, and describe the migration of the producing 
region in Section 4. Finally, based on comparing the butterfly diagram of prominence activities with the magnetic butterfly diagram, we discuss the 
origin of the unusual migration in the southern hemisphere and the anomalies of the magnetic field evolution.

\section{Observation}

The Nobeyama Radioheliograph (NoRH) is an interferometer dedicated to solar observations in the microwave \citep{Naka94}. 
The NoRH has been in operation for over 20 years since July 1992 and its rate of operation exceeds 98 \%. 
Although the main science target is the particle acceleration in a solar flare, the NoRH is one of the best instruments for monitoring of 
prominence activities because it can observe the Sun even on cloudy and rainy days and cover a wide field of view. 
The observing frequency of the data used in this paper is 17 GHz and the time resolution is three minutes. 
The spatial resolution (the size of the synthesized beam) has the seasonal and daily dependence and the average is about 10". 
The semi-automatic detection method of prominence activities for the NoRH data was developed by \citet{Shim06}. 
We applied the method to the data obtained from July 1992 to March 2013. The 1131 events are detected from more than 20 years of data, 
and a database was made that includes the time, size, position, radial velocity and MPEG movie of the prominence activities. 
The database is available at the website of the Nobeyama Solar Radio Observatory of the National Astronomical Observatory of Japan 
(http://solar.nro.nao.ac.jp/norh/html/prom\_html\_db/). The time variation of the statistical values (e.g. the number of events) 
during the solar cycle is discussed in this paper. The time resolution of such values is one year because the detecting efficiency 
of their method has a seasonal dependence due to the seasonal variation of the shape and size of the synthesized beam. 

To compare the migration of the producing regions of the prominence activities with photospheric magnetic field, 
we made the magnetic butterfly diagram from the synoptic Carrington rotation maps of magnetic field provided by 
the National Solar Observatory/Kitt Peak.

\section{Number variation of the prominence activities during Cycle 22--24}

\begin{figure}
  \begin{center}
    \FigureFile(80mm,80mm){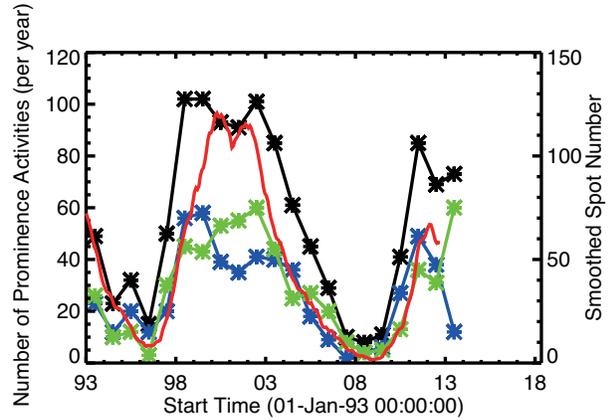}
  \end{center}
  \caption{
	The time variations of the number of the prominence activities and sunspots from 1993 to 2013. 
	The blue and green lines show the number variation of the prominence activities occurring in the northern and southern hemisphere, 
	respectively. The black line is the total. The red line indicates the smoothed sunspot number. 
	}
  \label{fig1}
\end{figure}

Figure \ref{fig1} shows the time variations of the numbers of the prominence activities and sunspots from 1993 to 2013. 
The number of the prominence activities is normalized in the observing days of each year. 
The value of 2013 has the large uncertainty because the database includes only three months' data. 
The sunspot number in the figure is the six-month running mean calculated from the data that are released by 
the Solar Influence Data Center of the Royal Observatory of Belgium \citep{Sidc92}.

Basically, the number variation of the prominence activities is similar to that of the sunspot, as already was mentioned by \citet{Shim06}. 
In Cycle 23, the peak in the northern hemisphere is earlier than that in the southern hemisphere, and the same tendency is seen in 
the time variation of the sunspot numbers.  When we pay attention to the timing of the minimum values between Cycles 22 and 23, 
the figure shows that the number of the prominence activities (in the northern/southern hemisphere) and 
the sunspots became the minimum in 1996. On the other hand, between Cycles 23 and 24, the timings of the minimum values are different. 
Although the number of prominence activities occurring in the northern hemisphere became the minimum in 2007, 
the data of the sunspots and the prominence activities in the southern hemisphere indicates that the minimum year is 2008. 
The number of prominence activities suggests that the solar cycle period of the southern hemisphere in Cycle 23  is longer than  
the period of the northern hemisphere, 11 years.

The yearly averages of the size and radial velocity also depend on the phase of the solar cycle, and become maximum value near the solar 
maximum. The time variation of the size shows the hemispheric asymmetry that is the same as the number variation, but the time variations of 
the velocity do not show any significant asymmetry of hemisphere in Cycles 23 and 24. 

\begin{figure*}[t]
  \begin{center}
    \FigureFile(160mm,100mm){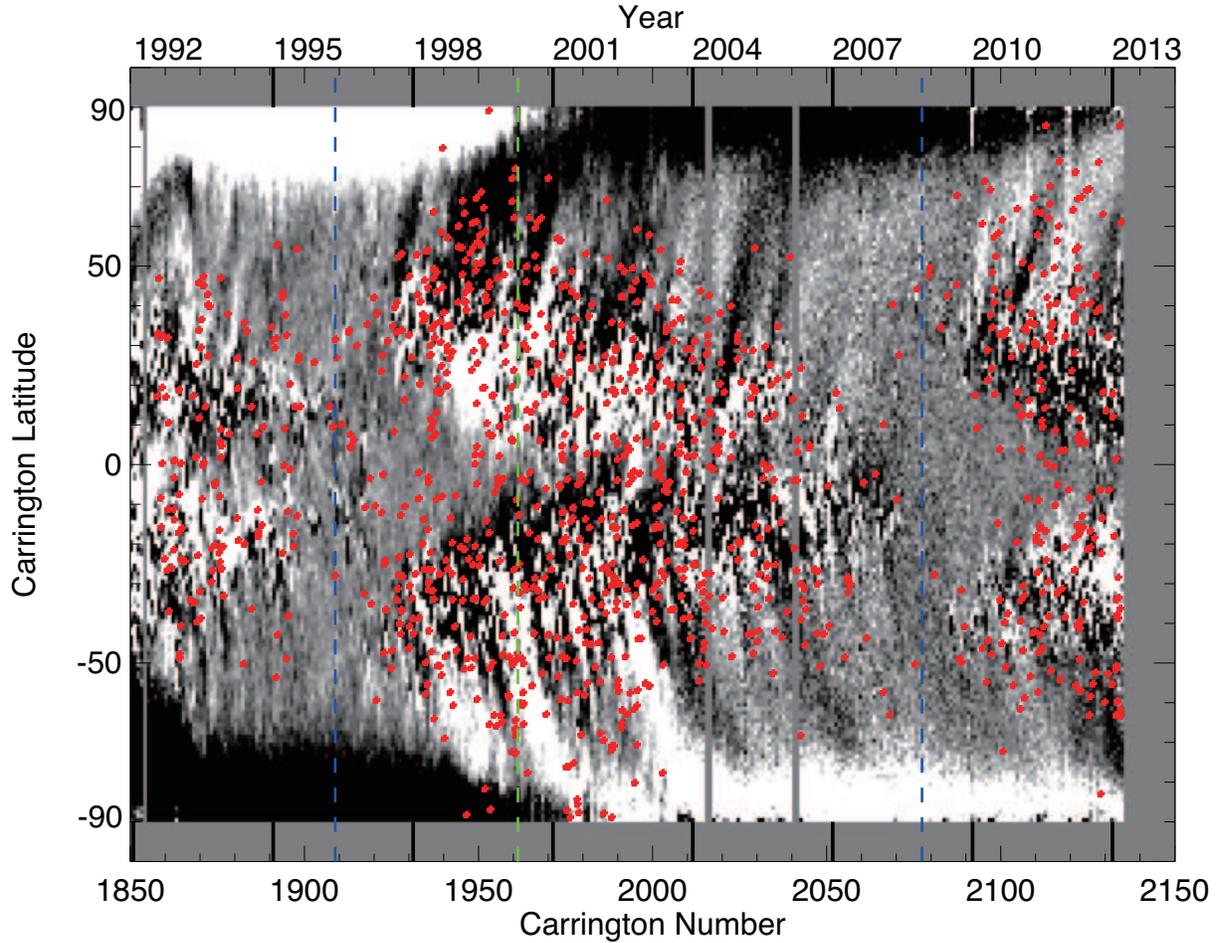}
  \end{center}
  \caption{
	The butterfly diagram of the prominence activities and the photospheric magnetic field. 
	The red dots indicate the dates and latitudes of the prominence activities and the grayscale shows the magnetic field distribution. 
	The blue and green dashed lines indicate the solar minimum and maximum.
	}
  \label{fig2}
\end{figure*}

\section{Migration of the producing region of the prominence activities}

Next, we consider the migration of the producing region of the prominence activities during the solar cycle. 
A prominence is the best proxy of the magnetic neutral line. 
Consequently, the migration of the producing region of prominence activities indicates the changing of the global distribution of 
the magnetic field in the Sun. Our analysis result of the prominence activities indicates that the migration in the southern hemisphere 
has been unusual from the solar maximum of Cycle 23. 
So, the prominence activities suggest that the anomalies of the global magnetic field distribution started at the solar maximum of Cycle 23.

In this section, we describe the details of the unusual migration in the southern hemisphere, 
but we start the section by describing the northern hemisphere in which the migration is similar to the previous cycles.

\begin{figure}
  \begin{center}
    \FigureFile(80mm,80mm){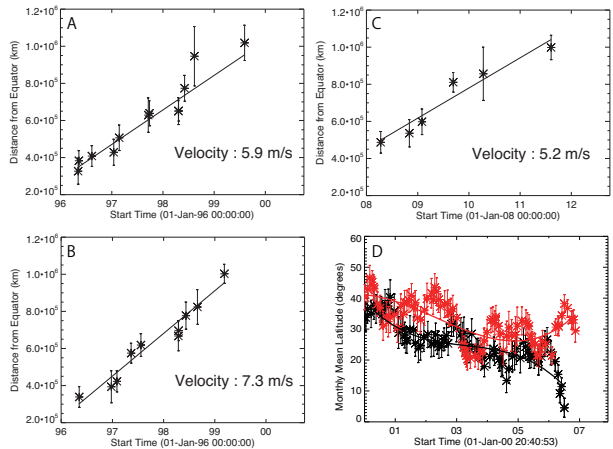}
  \end{center}
  \caption{
		Panels A, B, and C show the positions of the polemost prominence activities during the rising phase of the solar cycles. 
		A: Cycle 23/Northern hemisphere, B: Cycle 23/Southern hemisphere and C: Cycle 24/Northern hemisphere. The vertical axis indicates the 
		distance from the equator and covers from $\rm\pm$20 degrees to $\rm\pm$90 degrees latitude. 
		The error bars in the panels indicate the size of each prominence activity. 
		Panel D shows the monthly mean latitude of the prominence activities in the decay phase of Cycle 23. 
		The black and red asterisks indicate the monthly mean latitude in the northern and southern hemisphere, respectively. 
		The thick solid lines are the fitting results, and the error bars in Panel D indicate the monthly mean size of the prominence activities.	
		}
  \label{fig3}
\end{figure}

\subsection{Northern hemisphere}
 
Figure \ref{fig2} is the butterfly diagram of the prominence activities laid over the magnetic butterfly diagram, and covers over the 20 years that  
correspond to the period from the decay phase of Cycle 22 to the rising phase of Cycle 24. We concentrate on the northern hemisphere at this subsection.  
The production region of the prominence activities is expanding toward the pole in the rising phase of the solar cycle.  When the polemost 
prominence activity occurs near the pole around the solar maximum, the polarity reversal has taken place. The appearance region of dark filaments 
observed in H-alpha line also shows the similar poleward expansion and the relation with the polarity reversal 
\citep{Wald73,Topk82,Mour94,Maka01,Li08}, and \citet{Gopa03} also reported the relation between the prominence activities and the polarity 
reversal. The velocity of the poleward expansion can be derived by tracing the polemost prominence activities. Panels A and C in Figure \ref{fig3} 
are the results of tracing during the rising phase of Cycles 23 and 24, and show that the expansion velocities in the northern hemisphere are 
5.9 m/sec at Cycle 23 and 5.2 m/sec at Cycle 24. 
\citet{Hath10} derived the latitudinal profile of the meridional flow speed from the magnetic features 
observed by the Michelson Doppler Imager (MDI) aboard the Solar and Heliospheric Observatory (SOHO) from May 1996 to June 2009. The 
expansion velocity that is derived from the prominence activities  corresponds to the meridional flow speed around 60 degrees latitude. The 
coincidence suggests that the poleward migration of the polemost prominence activities is an indicator of the meridional flow. Our result is slower 
than the expansion velocity ($\rm\sim$ 10 m/sec) derived from the dark filaments observed in Cycle 19 \citep{Topk82}. The disagreement might 
show that the meridional flow speed in Cycle 23--24 is slower than that in the previous cycles.

After the solar maximum, the producing region of the prominence activities is shrinking toward to the equator. The variation of the producing 
region agrees with the variation of the active region filaments rather than all dark filaments \citep{Mour94}. This suggests that most of the prominence 
activities in the decay phase of a solar cycle are produced by the prominences in active regions.  \citet{Li10} investigated the appearance region of 
the dark filaments in Cycle 16--21, and suggests that a third-order polynomial curve could give a satisfactory fit to the monthly mean latitude of 
dark filaments. We also applied the fitting to our database, and the thick black line of Panel D in Figure \ref{fig3} is the fitting result. The fitted mean 
latitude in the northern hemisphere at Cycle 23 is similar to that at the previous cycles given in Figure 4 of \citet{Li10}, and the migration velocity is 
also similar to his result.
  
According to the observing facts (an 11-year cycle period, the velocity of expanding migration, and the migration of mean latitude at the decay 
phase), we can say that the migration of the producing region of the prominence activities in the northern hemisphere at Cycle 23-24 is similar to 
the previous cycles. When the butterfly diagram of prominence activities that is shifted 11 years is overlaid on Figure \ref{fig2} in the northern hemisphere 
(the overlaid figure is not included in the paper), the prominence activities that occurred after the solar maximum of Cycle 23 trace the 
butterfly diagram of the prominence activities from the decay phase of Cycle 22 to the rising phase of Cycle 23. The fact also suggests that the 
northern hemisphere at Cycle 23--24 is normal from the point of view of prominence activities. 

\subsection{Southern hemisphere}

At the rising phase of Cycle 23, the producing region of the prominence activities is expanding toward the pole in the southern hemisphere.  The 
expansion velocity is 7.3 m/sec (Panel B of Figure \ref{fig2}) and is faster than that in the northern hemisphere. 
The same trend of the hemispheric asymmetry is reported by \citet{Hath10}, 
and it also suggests that the poleward migration of polemost prominence activities indicates the meridional flow.

After the solar maximum of Cycle 23, the migration of the producing region of the prominence activities became unusual. 
The thick red line in Panel D of Figure \ref{fig3} shows the fitting result of the monthly mean latitude 
in the southern hemisphere at the decay phase of Cycle 23. Although the 
difference of the mean latitude migration in the northern and southern hemisphere at the previous cycles is small \citep{Li10}, the migration of the 
southern hemisphere is significantly different from that of the northern hemisphere at Cycle 23 and the previous cycles. 
The monthly mean latitude in the southern hemisphere did not decrease quickly after the solar maximum as in  the northern hemisphere, 
and it stayed over -30 degree latitude after 2006. 
Figure 2 shows that the prominence activities occurred at over -50 degree latitude after 2006. Surprisingly, 
some prominence activities occurred at over -60 degrees latitude even in 2008, the solar minimum year. So, such high latitude prominence activities 
caused the unusual migration in the decay phase at the southern hemisphere. 

The rising phase of Cycle 24 began in 2009. Although we can see the prominence activities associated with the active regions from late 2009, 
the expansion of the producing region in the southern hemisphere is not seen in the butterfly diagram of prominence activities. 
Only the two prominence activities occurred at over -65 degrees latitude after 2009.

\begin{figure*}[t]
  \begin{center}
    \FigureFile(160mm,100mm){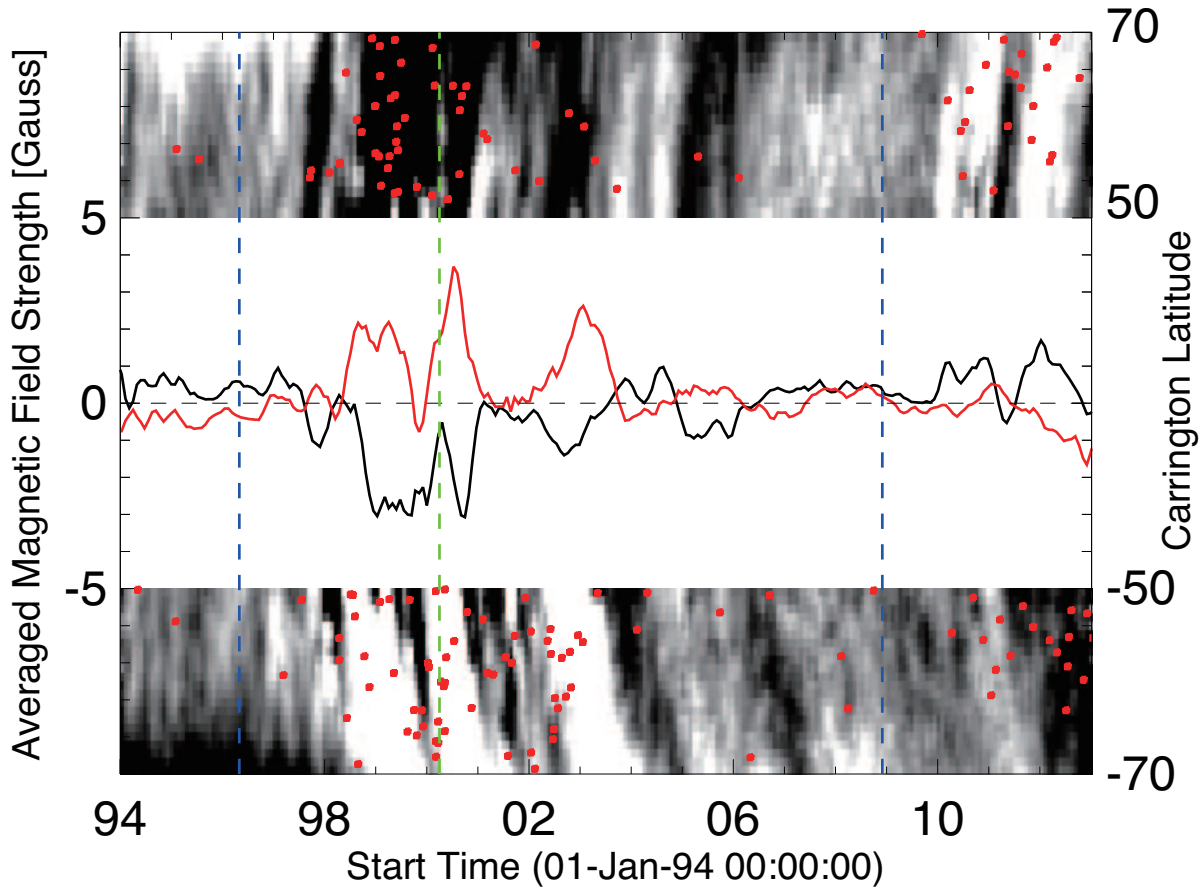}
  \end{center}
  \caption{
		The time variation of the magnetic field distribution in the high-latitude regions ($\rm\pm$50--70 degrees). The grayscale of the upper 
		and lower parts in the figure shows the magnetic butterfly diagram of the high-latitude region in the northern and 
		southern hemisphere, respectively. The red dots on the parts indicate the dates and latitudes of the prominence activities. 
		The black and red lines in the middle part show the time variation of the averaged magnetic field strength 
		in the high-latitude regions of the northern and southern hemispheres. 
		The blue and green dashed lines indicate the solar minimum and maximum.
		}
  \label{fig4}
\end{figure*}

\section{Discussion}

In the previous section, we described the migration of the producing region of the prominence activities using the butterfly diagram of prominence 
activities, and showed that the unusual migration in the southern hemisphere started from the solar maximum of Cycle 23. It is clear that the unusual 
migration was caused by the anomalous prominence activities in the high-latitude region (over -50 degrees). To understand what produces the 
anomalous prominence activities and the anomalies of the global magnetic field distribution, we investigated the relation between the high-latitude 
prominence activities and the photospheric magnetic field.

Figure \ref{fig4} shows the time variation of the magnetic field at the high-latitude region ($\rm\pm$50--70 degrees). The magnetic butterfly 
diagram of the high-latitude region (the upper and lower parts of Figure \ref{fig4}) clearly shows the \textquotedblleft rush to the pole" of magnetic elements, and 
suggests that the prominence activities are closely related to the rushes to the pole. First, we turn our attention to the late decay phase of Cycle 23 
(2005$\rm\sim$). The last rush to the pole of Cycle 23 in the northern hemisphere started in 2005 and the magnetic polarity of the rush has the 
same polarity as the north polar region. The similarity of the polarity does not destabilize the polar crown prominences; therefore the prominence 
activities did not occur in the high-latitude region. After the last rush to the pole in the northern hemisphere, the polarity in the high-latitude region 
became plus and stabilized (see the middle part of Figure \ref{fig4}). The properties of the magnetic field might restrain the prominence activities in 
the high-latitude region of the northern hemisphere till starting the rising phase of Cycle 24. In the southern hemisphere, the last rush to the pole 
started in 2006, one year later from the last rush in the northern hemisphere. Furthermore, the magnetic polarity of the rush has the anti-polarity of 
the south polar region, hence the rush destabilized the prominences. Since the end of the last rush to the pole in the southern hemisphere (after 2007), 
the polarity of the high-latitude region has not been stable. Such magnetic complexity in the southern 
high-latitude region induces the prominence activities, and caused the unusual migration in the decay phase of Cycle 23. Considering that the rush 
to the pole is made of sunspots, the hemispheric asymmetry of the prominence activities indicates that the southern hemisphere was more 
active in the late decay phase of Cycle 23. The inference is consistent with the magnetic field observation reported by \citet{Petr12}. Additionally, 
we think that the prominence activities in the high-latitude region are related to the polar magnetic field. Figure \ref{fig2} shows that the polar magnetic 
field expanded from the pole until $\rm\pm$65 degrees latitude in the decay phase of Cycle 22. Such a single polarity region might restrain 
prominence activities. On the other hand, the polar magnetic field in the decay phase of Cycle 23 did not expand over 
$\rm\pm$75 degrees. The long-term magnetic butterfly diagram also indicates that the strength and size of the polar magnetic field in the decay 
phase of Cycle 23 are significantly weaker and smaller than those of the previous cycles (e.g. \cite{Wang09,Petr12,Gopa12}). The weak polar 
magnetic field might make the complexity in the high-latitude region and induce the prominence activities. 

After starting the rising phase of a solar cycle, the expanding of the producing region of the prominence activities appears as usual. In the 
northern hemisphere, the expansion appeared as usual, and the polar reversal might be finished in 2013 because the polemost prominence activity 
had already reached around the pole. As we have already mentioned, in the southern hemisphere, the prominence activities over -75 degrees latitude are very 
rare and the expansion cannot be seen significantly though the prominence activities with the active regions of Cycle 24 occurred at the same time 
as in the northern hemisphere (Figure \ref{fig2}). The lower part of Figure \ref{fig4} shows that the first strong rush to the pole in the southern 
hemisphere has a polarity that is the same as that of the polar field in the solar minimum of Cycle 23. Because such a similarity of the 
polarity restrains prominence activities, the prominence activities did not occur over -65 degrees latitude until 2011. In 2011, the rush to the pole 
with the anti-polarity of the south polar region appeared. However, the magnetic field strength of the rush is weak, less than 1 Gauss and was 
not able to produce the prominence activities. The magnetic field strength of the rush with negative polarity in the southern hemisphere is 
increasing and the absolute strength became similar to that of the rush in the northern hemisphere in late 2012. Based on the observing facts, we 
predict that the prominence activities occur more frequently in the high-latitude region from 2013. 
This means that the polarity reversal in the southern hemisphere is delayed from the northern hemisphere, as suggested by some authors 
(e.g. \cite{Shio12,Sval13}). 
Finally, we mention that the magnetic field strength of the rushes in Cycle 24 is significantly smaller than that of Cycle 23 in both of the 
hemispheres, consistent with our finding that the number of the prominence activities in the high-latitude region at Cycle 24 is smaller than that at Cycle 23.
However, the influence does not appear in the total number of the prominence activities, as shown in Figure 1, 
because most of the prominence activities occur in the active region.

In the paper, we described the unusual migration of the producing region of the prominence activities in the southern hemisphere and interpreted 
the anomalies from the distribution of the photospheric magnetic field. However, the origin of the anomalies is hidden under the photoshere, and 
we need the progress of the solar dynamo studies to fully understand the anomalies. We show that a polemost prominence activity is a good 
indicator of the meridional flow. The tracing of the polemost filament or prominence in the previous cycles might be one of the keys to 
understand the meridional flow and the solar activities.

\bigskip

The authors thank to all current and former staffs of the Nobeyama Solar Radio Observatory for their effort to ensure the stable operation of 
the NoRH for over 20 years. NSO/Kitt Peak data used here are produced cooperatively by NSF/NOAO, NASA/GSFC, and NOAA/SEL. 
This work was carried out on the Solar Data Analysis System (SDAS) operated by the Astronomy Data Center in cooperation with the Nobeyama 
Solar Radio Observatory of the National Astronomical Observatory of Japan.


\end{document}